\documentclass[twocolumn,preprintnumbers,12pts,showpacs,showkeys]{revtex4}
\usepackage{graphicx}
\usepackage{amsmath,amssymb}
\usepackage{epstopdf}
\usepackage{float}
\usepackage{color}
\usepackage{hyperref}
\graphicspath{images/}
\begin{document}
\title{Spatio-temporal of Dirac and Klein Gordon particles in a one-dimensional box}
\author{$^{1}$Kamran Ullah, $^{2}$Mustafa Khan, $^{3}$Jameel Hussain} \email{kamran@phys.qau.edu.pk$^{1}$}
\affiliation{$^{1,2}$Department of Physics, Quaid-i-Azam University, 45320 \ Islamabad, Pakistan}
\email{mkhan.quantum@gmail.com$^{2}$}
\affiliation{$^{3}$Department of Electronics, Quaid-i-Azam University, 45320 \ Islamabad, Pakistan.}
\begin{abstract}
Present work is devoted to studying the spatio-temporal of the Dirac and the Klein Gordon (KG) particles confined in a one-dimensional box. We discuss the quantum carpets and revivals time for each particle. Moreover, we explain that the relativistic phenomena not only decrease the density of the corresponding carpets but also affect the revivals time of each particle within a box. Further, we impose different limiting conditions on the obtained Dirac energy solution and explain the well-known slight-relativistic and non-relativistic quantum carpets, respectively. In addition, we extend our study to explain the quantum carpets of the KG particle at different energy states and compare its revivals time with the Dirac particle.
\end{abstract}



\date{\today}
\maketitle
\section{Introduction}\label{sec:intro}
Spatio-temporal patterns of the quantum mechanical probability density is a manifestation of quantum interference between different eigenmodes of the system which exhibit quantum carpets. In a bound system, the spatio-temporal evolution of a wave packet has been not only studied in a simple systems \cite{1,a}, but also discussed in a complex system \cite{2}. It is well known that the quantum carpets can be analyzed in those system where the energy spectrum has a quadratic dependency, for example, the near-field diffraction of Bose-Einstein condensates confined in a one-dimensional hard wall potential \cite{3}. In addition, the quantum carpets have also been studied in a fractional dimensional space \cite{4,5}, and in continuous time random walk \cite{6}.

Among these lines, particle in a box provides a playground for the study of the quantum revivals and spatio-temporal evolution of a particle by which we explain the quantum carpets. Different approaches were used to realize the quantum carpets and revivals both in the non-relativistic and slight relativistic regimes, names of few, Wigner functions technique \cite{7}, Green's fuction \cite {8}, and probability averaging \cite{8a}, and so on. In addition, it has been studied that the bound localized Dirac particle exhibits revival of the Zitterbewegung oscillations amplitude \cite{9}, which decreases from period to period. Furthermore, the quantum carpets for a relativistic Dirac particle confined in a circular ring of radius R has been discussed in \cite{10}. 

Besides, some experimental works have also been done on the quantum Talbot effects similar to quantum carpets and revivals. The most interesting one is the quantum Talbot effects with single photons and entangled photon pairs which has been recently studied in \cite{11}. In addition, the other experimental studies on the quantum carpets and revivals are Rydberg atom \cite{12}, and semiconductor well \cite{13}, respectively.

In this work, we discuss the quantum carpets and revivals for Dirac and Klein Gordon particles confined in a one-dimensional box. Further, we explain that both particles completely reconstructs their original shape in a certain period time, called revival time. In addition, we also explain that the relativistic quantum carpets have a different revival time from both non-relativistic and slight relativistic quantum carpets. It is because the relativity of many revivals are washed out, as a result, we have less denser relativistic quantum carpets.

It is well known that the particle in a box is a perfect example of a bound system and such a system is suitable to discuss the quantum carpets and revivals. Let us consider a particle of mass m confined in a one dimensional box through a potential 
\begin{equation}\label{eq:cond}
 V(z)=\begin{cases}
0 & \text{if $0 < z < L$}\\
\infty & \text{otherwise}.
\end{cases}
\end{equation} 
 The dyanmaics of Eq.~\eqref{eq:cond} is governed by the time-independent Schr\"{o}dinger wave equation i.e.,
\begin{equation}\label{eq:schrodinger}
-\frac{\hbar^{2}}{2m}\frac{\text{d}^2\psi(z)}{\text{d}z^2}=E\psi(z).
\end{equation}
 Whose initial wave solution at time $t=0$ is
\begin{equation}\label{eq:wave}
\Psi_m(z,0)=\sum_{m=1}^{\infty}C_m\psi_m(z), 
\end{equation}
 where $$\psi_{m}(z)=\sqrt{\frac{2}{L}}\sin(\frac{m\pi z}{L})$$ is the general solution of time-independent Schr\"{o}dinger wave equation, and $C_m$ is the expansion coefficient which can be calculated from the relation $C_{m}=\int_{0}^{L}\Psi_{m}^{\star}(z,0)\psi_{m}(z)dz$ \cite{2}. Here, $$\psi_{m}(z,0)=\frac{1}{(\pi\Delta^{2})^{1/4}}\exp[-(z-z_{0})^2/2\Delta^{2}]\exp[ip_{0}(z-z_0)/\hbar]$$ is the Gaussian wave packet initially prepared at time t=0, of width $\Delta=10^{-2}$ centered at $z_{0}=L/2=0.5$. Inserting $\psi_{m}(z,0)$ and $\psi_{m}(z)$ in Eq.~\eqref{eq:wave}, the eigen wavefunction for the non-relativistic particle confined in a one dimensional box at time $t=0$ can be described as
\begin{equation}\label{eq:sol}
\begin{split}
\Psi_{m}(z, 0)=&\sqrt{1/\pi\Delta L}\sum_{m=1}^{\infty}\exp[-\frac{\pi^{2}\Delta^{2}}{2L^{2}}m^{2}]\sin(\frac{m\pi z}{L})\times\\&\sin(\frac{m\pi z_{0}}{L}).
\end{split}
\end{equation} 
Here, $z_{0}$ represents the position of the initial wave packet and L is the length of the box. Eq.~\eqref{eq:sol} represents the spatial part of the non-relativistic particle within a 1-D box. For the space-time dynamics, we multiply Eq.~\eqref{eq:sol} by $\exp[-iE_{n}t/\hbar]$, where $E_{n}$ is the $n^{th}$ eigen energy state of the particle. The quantum carpets for a non-relativistic particle confined in a one-dimensional box, whose probability density ($\rho(z,t)=\Psi_{m}^{\star}(z,t)\Psi_{m}(z,t)$) after time t is given by
\begin{equation}\label{eq:probability}
\begin{split}
\rho(z, t)=&1/\pi\Delta L\sum_{m=1}^{\infty}\sum_{n=1}^{\infty}\exp[-\frac{\pi^{2}\Delta^{2}}{2L^{2}}(m^{2}+n^{2})]\sin(\frac{m\pi z}{L})\\&\times\sin(\frac{n\pi z}{L})\sin(\frac{m\pi z_{0}}{L})\sin(\frac{n\pi z_{0}}{L})\\&\times\exp[-i\frac{(E_{n}-E_{m})t}{\hbar}].
\end{split}
\end{equation}
In Eq.~\eqref{eq:probability}, t is the evolution time for the non-relativistic particle and the dummy indices in summation represents the quantum numbers associated with different eigen-energy states.
The rest part of the paper is organized as:\\
In Sec. II, we study the spatio-temporal evolution of the Dirac particle confined in a one-dimensional box and discuss its limiting cases both in the non-relativistic and slight relativistic regimes. In Sec. III, we explain the spatio-temporal evolution of the KG particle in a 1-D box and discuss how its quantum carpets are different from the Dirac particle. In Sec. IV, we conclude our results.
\section{ Spatio-temporal of Dirac particle} 
In this section, we discuss the spatio-temporal of Dirac particle confined within a one-dimensional box. We consider that the potential of the box is the same as given in Sec. I. Therefore, the eigen energy equation for the spatial part of the Dirac particle confined in a one dimensional box can be represented as \cite{16,17,18}. 
\begin{equation}\label{eq:dirac}
H\phi(z)=(-i\hbar c\alpha_{z}\frac{d}{d z}+m_{0}c^{2}\beta)\phi(z)=E\phi(z),
\end{equation}
where $\alpha_{z}$ and $\beta$ are the Pauli spin matrices written as: $\alpha_{z}=
\begin{pmatrix}
	0 & \sigma_{z}\\
	\sigma_{z} & 0
\end{pmatrix}$, and $\beta=\begin{pmatrix}
	I & 0 \\
	0 & -I
\end{pmatrix}$,
 where $\sigma_{z}$ is $2\times2$ Pauli-spin matrix and I is the $2\times2$ unit matrix, respectively, which satisfy $\alpha_{z}\beta+\beta\alpha_{z}=0$ with $\alpha_{z}^{2}=\beta^{2}=1$.
Using boundary condition $\pm(-i)\beta\alpha_{z}\phi=\phi$ at $z=L$ (MIT bag model)~\cite{mit1,mit2, mit3}. The solution of Dirac equation within the box is
\begin{equation}\label{eq:wfn}
\phi(z)=B\exp{i \delta/2}\begin{pmatrix} 2\cos{(kz-\delta/2)\chi}\\
2iP\sin{(kz-\delta/2)\sigma_{z}\chi}
\end{pmatrix}.
\end{equation}
 Where B is the normalization constant and\ $$\delta=\arctan(\frac{2P}{P^2-1}).$$ Here, P is dimensionless momentum\ $$P=\frac{\hbar k c}{E+m_{0}c^2}.$$ Where E is the energy of the particle and k is the wave vector. Further, $\chi$ is an arbitrary two-component normalized spinor which satisfy the condition $\chi^\dagger\chi=1$ with $\chi=\begin{pmatrix}1\\0\end{pmatrix}$ for spin "up" and $\chi=\begin{pmatrix}0\\1\end{pmatrix}$ for spin "down". To write the most general form of the eigen wavefunction for a Dirac particle confined in a one dimensional box, vanishing at $\mid z\mid\rightarrow\infty$ . Again recall the boundary condition $\pm(-i)\beta\alpha_{z}\phi=\phi$ with $+$ and $-$ corresponding to z=0 and z=L, respectively. Invoke the arbitrariness of $\chi$, and use Dirac spinors, the wavefunction for boundary condition z=L is obtained
\begin{equation}
\begin{split}
\phi_{m}(z)=&\sum_{m=1}^{\infty}B_{m}[(1+P_{m})\exp[ik_{m}z]+(1-P_{m})\exp[i(\delta_{m}\\&-k_{m}z)].
\end{split}
\end{equation} 
Here, $P_{m}=\hbar k_{m}c/(E_{m}+m_{0}c^{2})$, $k_{m}=m\pi/L$ and $\delta_{m}=\arctan[2P_{m}/(P^{2}_{m}-1)]$ are respectively momentum, wave number and phase of the Dirac particle. In addition, $B_{m}$ represents the probability amplitudes which can be evaluated from the initial condition as discussed in Sec. I. Therefore, the complete eigenfunction of the particle at t=0 can be evaluated as
\begin{equation}
\begin{split}
\Phi_{m}(z,0)=&8\pi\Delta\sum_{m=1}^{\infty}([\cos(k_{m}z-\frac{\delta_{m}}{2})-iP_{m}\sin(k_{m}z-\frac{\delta_{m}}{2}\\&)][\cos(k_{m}z_{0}-\frac{\delta_{m}}{2})+iP_{m}\sin(k_{m}z_{0}-\frac{\delta_{m}}{2})])\\&\times\exp[-\Delta^{2}k_{m}^{2}],
\end{split}
\end{equation}

 where $\Delta$ represents the width of the initial Gaussian wave packet. The complete wave function of the particle after time t is $\Psi_{m}(z,t)=\Phi_{m}(z,0)\exp[\frac{-iE_{m}t}{\hbar}]$. This leads the probability density $\wp(z, t)=\Psi_{m}^{\star}(z,t)\Psi_{m}(z,t))=\sum_{m=1}^{\infty}\sum_{n=1}^{\infty}\Phi_{m}(z,0)\exp[\frac{iE_{m}t}{\hbar}].\Phi_{n}(z,0)\exp[\frac{-iE_{n}t}{\hbar}]$ which can be expressed as
\begin{equation}\label{eq:prob}
\begin{split}
\wp(z, t)=&64\pi^{2}\Delta^{2}\sum_{m=1}^{\infty}\sum_{n=1}^{\infty}\exp[-\Delta^{2}(k_{m}^{2}+k_{n}^{2})]([\cos(k_{m}z-\\&\frac{\delta_{m}}{2})-iP_{m}\sin(k_{m}z-\frac{\delta_{m}}{2})][\cos(k_{m}z_{0}-\frac{\delta_{m}}{2})+\\&iP_{m}\sin(k_{m}z_{0}-\frac{\delta_{m}}{2})][\cos(k_{n}z-\frac{\delta_{n}}{2})-iP_{n}\\&\sin(k_{n}z-\frac{\delta_{n}}{2})][\cos(k_{n}z_{0}-\frac{\delta_{n}}{2})+iP_{n}\sin(k_{n}z_{0}\\&-\frac{\delta_{n}}{2})])\times\exp[-i\frac{(E_{n}-E_{m})t}{\hbar}].
\end{split}
\end{equation}
In Eq.~\eqref{eq:prob}, $E_{n}=m_{0}c^{2}\sqrt{1+(2nq)^{2}}$ is the $n^{th}$ state energy and $m_{0}$ is the rest mass of the spin-1/2 particle. Moreover, $q=2\pi\hbar/4m_{0}cL=\lambda_{c}/4L$, where $\lambda_{c}=2\pi\hbar/m_{0}c$ is called Compton wavelength of the particle \cite{9}.\\
Next, we discuss the special cases of Eq.~\eqref{eq:prob} for different regimes: non-relativistic and slight relativistic, respectively. For non-relativistic case, we consider $\lambda_{c}\ll L$ and the relativistic parameter takes the value $q=10^{-6}$. Under these limiting conditions, the relativistic energy and momentum of the particle is reduced to a non-relativistic energy and momentum i.e.,
\begin{figure}[ht]
	\centering
	\includegraphics[height=6.5cm,width=4cm]{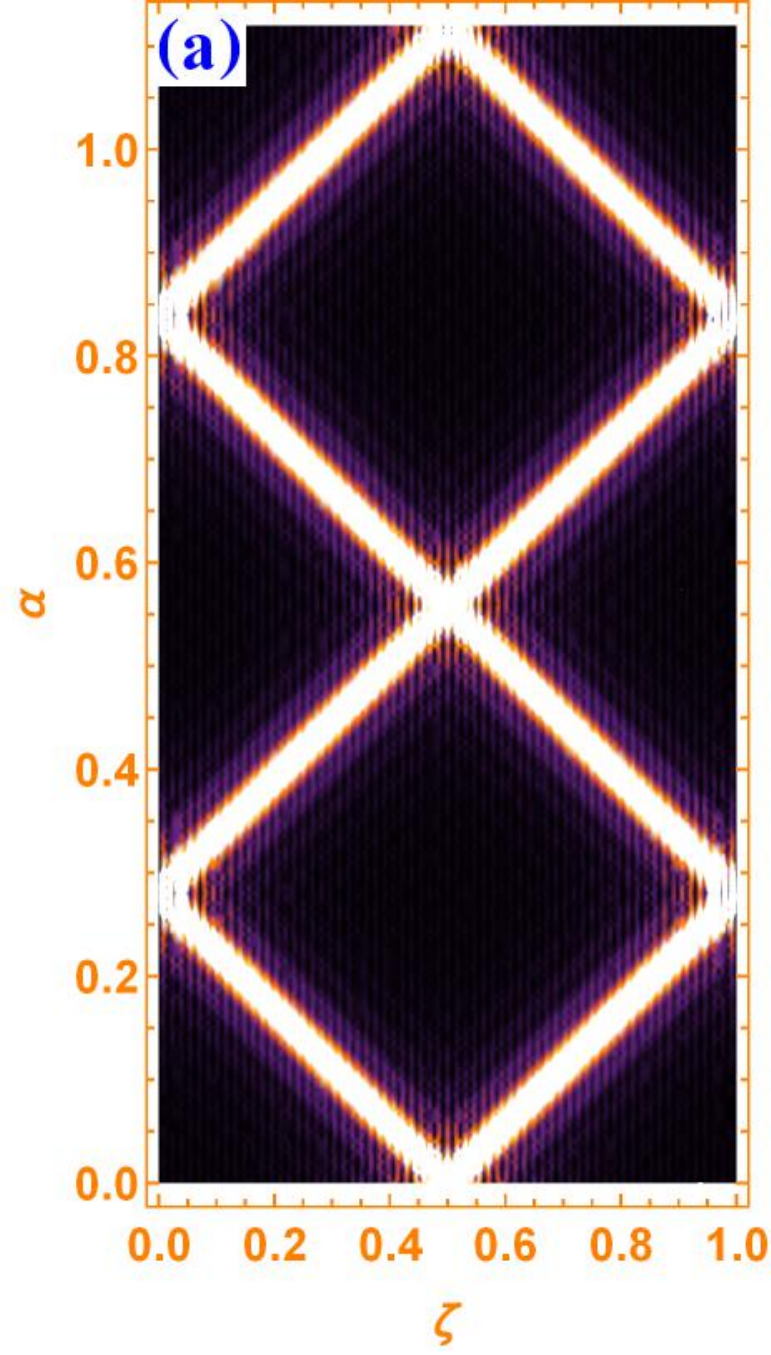}
	\includegraphics[height=6.5cm,width=4cm]{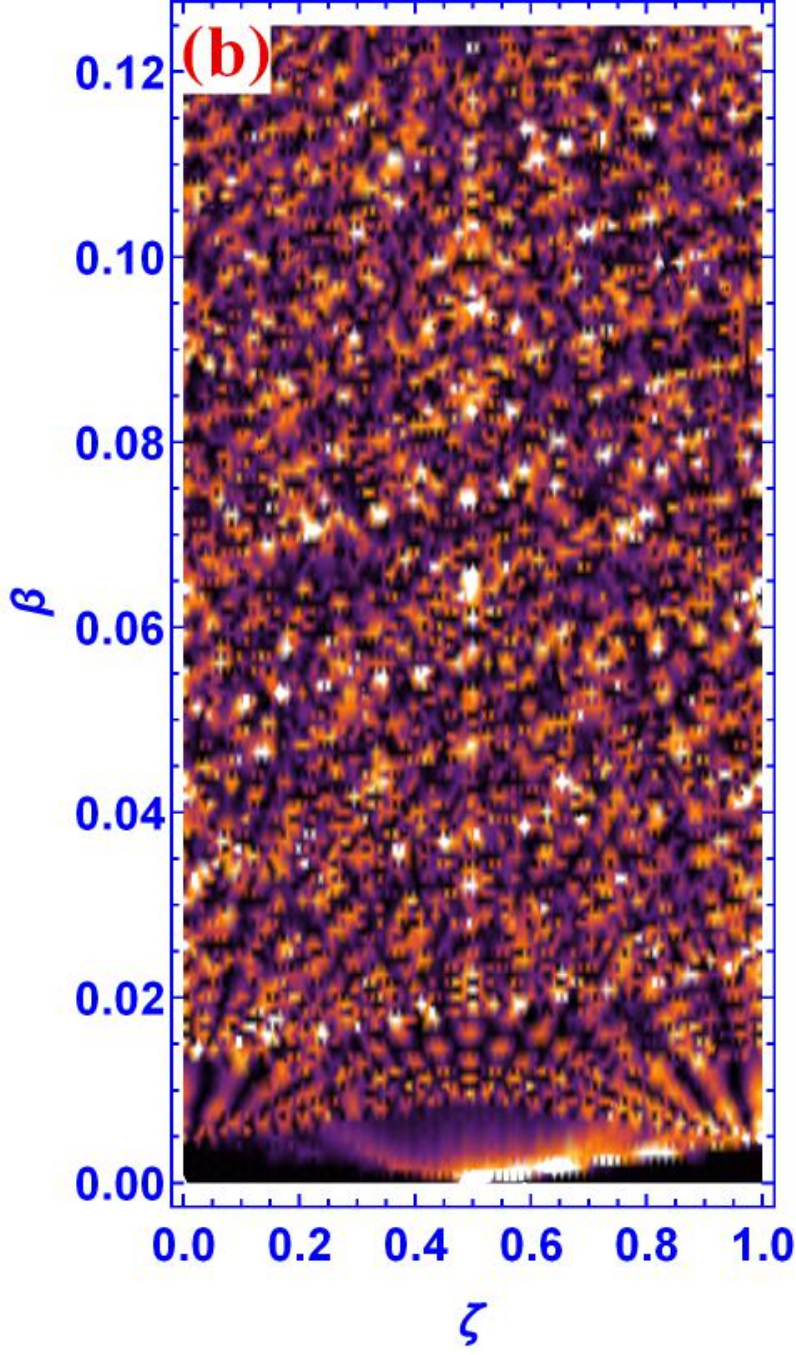}
	\includegraphics[height=6.5cm,width=4cm]{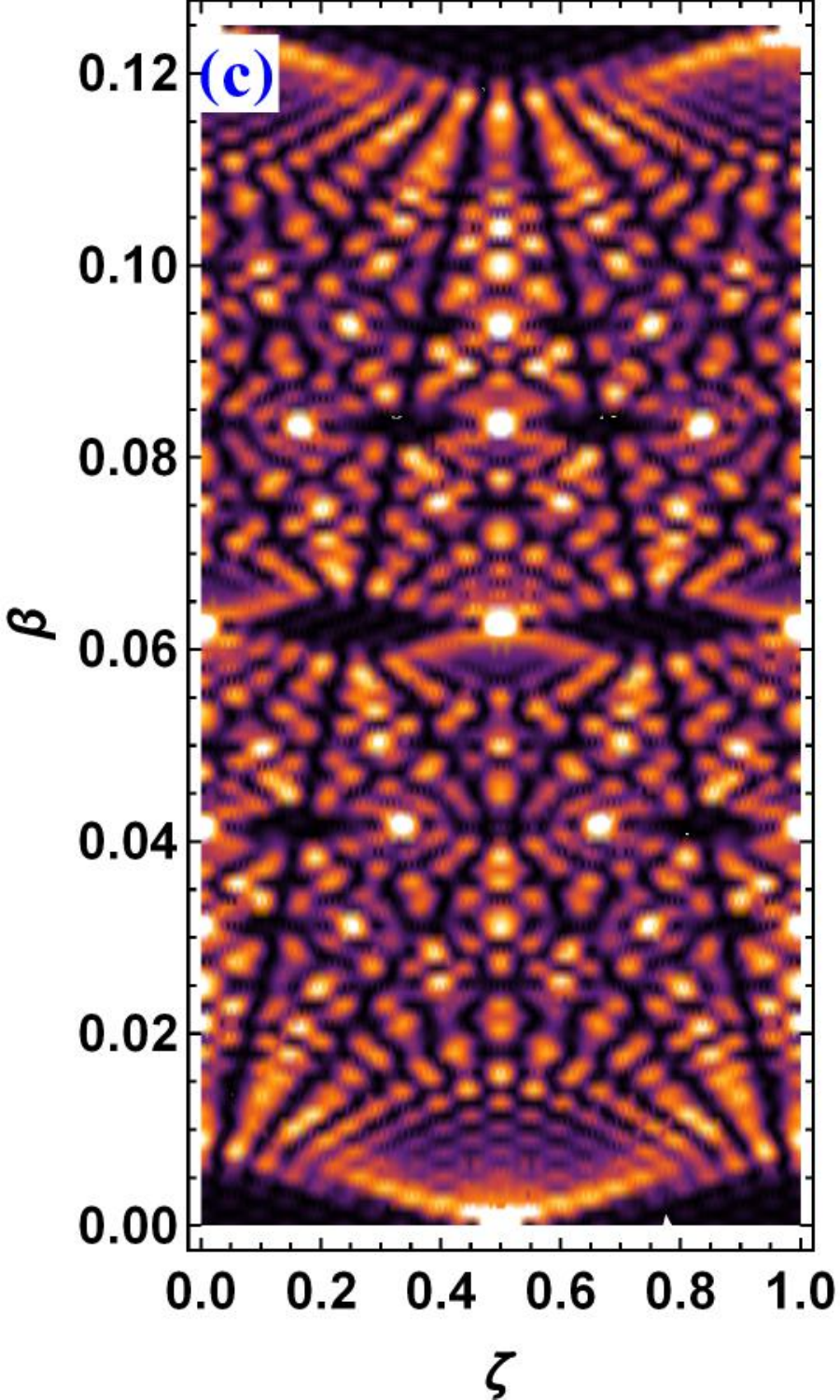}
	\includegraphics[height=6.5cm,width=4cm]{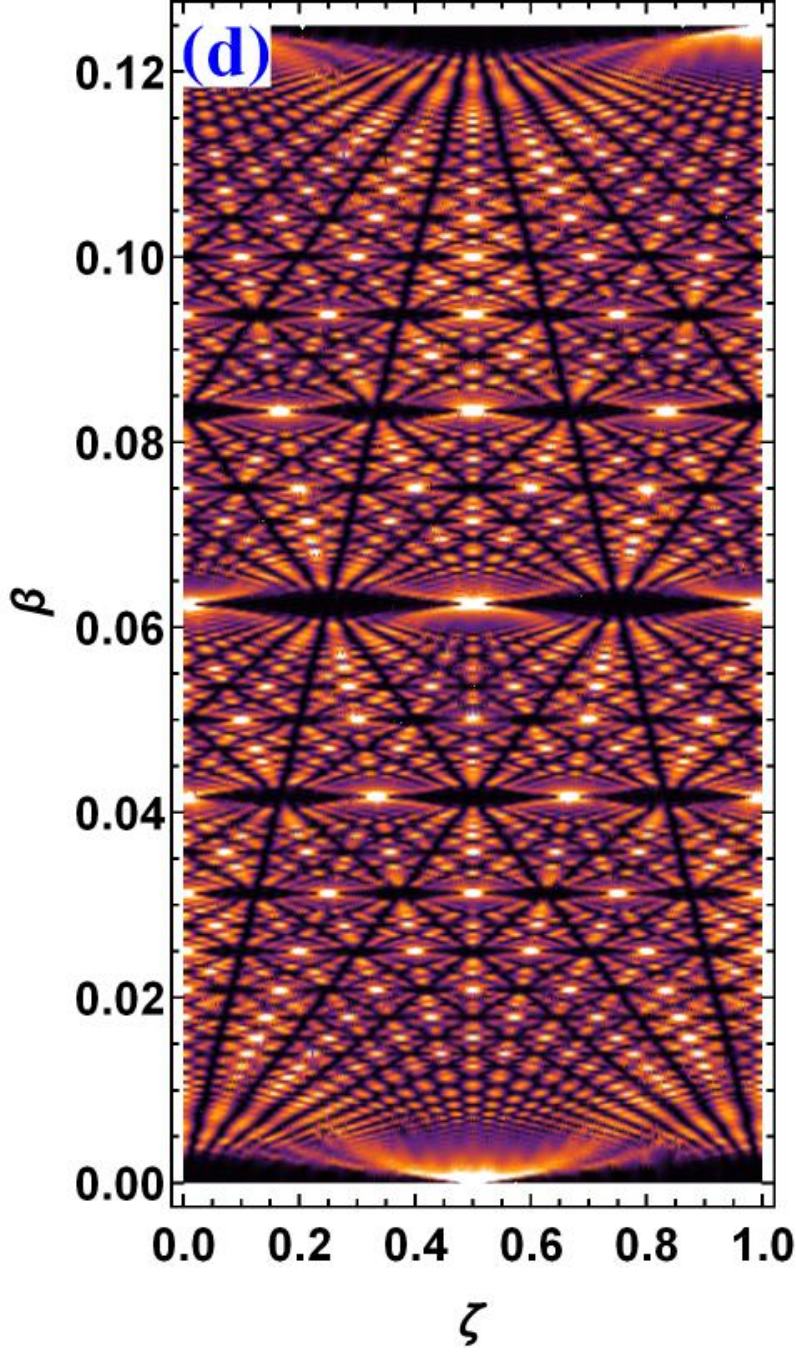}
	\caption{We plot the spatio-temporal of the Dirac particle in a one-dimensional box for different cases (a) relativistic (b) slight relativistic and (c,d) non-relativistic versus scaled length $\zeta=z/L$ for (a-d) and time $\alpha=t/T^{rev}_{Dirac}$ for (a) whereas $\beta=t/T^{rev}_{Sch}$ is the scaled time in (b-d). We take the width of the Gaussian wave packet $\Delta=10^{-2}$ for all cases.}
\end{figure}
\begin{equation}\label{eq:cases}
\begin{split}
E_{n}\cong\frac{2\pi n^{2}}{T^{rev}_{Sch}}, \\ P_{n}=nq.
\end{split} 
\end{equation}
Where $T^{rev}_{Sch}$ represent the revival time of the non-relativistic particle as earlier discussed in \cite{2}. Next, for the slight relativistic case, we take $\lambda_{c}<L$ and $q=10^{-2}$, therefore, Eq.~\eqref{eq:cases} can be modified as
\begin{equation}
\begin{split}\label{eq:modified}
E_{n}\cong\frac{2\pi n^{2}}{T^{rev}_{Sch}}-\frac{2\pi n^{4}q^{2}}{T^{rev}_{Sch}}, \\
P_{n}=nq-n^{3}q^{3}. 
\end{split}
\end{equation}
In Eq.~\eqref{eq:modified}, the $2^{nd}$ term is appeared due to correction in the slight relativistic energy. In the presence of these terms, the intermode traces the wave packet, as a result, we have no prominent reconstruction in the early evolution of the wave packet.

In Fig. 1(a), we plot Eq.~\eqref{eq:prob} for relativistic quantum carpets against the normalized length $z/L$ and rescaled time $t/T^{rev}_{Dirac}$. Initially, the particle is placed at the center of the box, which starts its early evolution from 0 and reconstruct itself at time $T^{rev}_{Dirac}$, where $T^{rev}_{Dirac}=\pi\hbar\sqrt{1+4n_{0}^{2}q^{2}}/(2m_{0}c^{2}q^{2}n_{0})$ with $q=1$. In addition, it is clear from Fig. 1(a) at time $ t=T^{rev}_{Dirac}/2$ (half of revival time), we observe symmetric revival of the initial wave packet due to existence of symmetry in the box. Noted that, in this particular plot, we take  $m_{min}=65$ and $m_{max}=95$. Hence, Fig. 1(a) reveals that the observed pattern demonstrates the dynamics of the fast-moving wave packets which collides with the wall of the box and reshape itself in a certain time due to the quantum interference between the eigen modes of the system.\\
Next, Fig. 1(b) displays the spatio-temporal of the slight-relativistic case against the normalized length $z/L$ versus normalized time $\beta=t/T^{rev}_{Sch}$. It is quite clear from Fig. 1(b) that the initial wave packet is unable to reconstruct itself in specific time because the higher order correction in energy ($n^{4}$) and in momentum ($n^{3}$), respectively, exist in Eq.~\eqref{eq:modified}. In this particular case, we take $m_{min}=2$ and $m_{max}=50$.\\ 
Similarly, in Figs. 1(c,d), we show the non relativistic quantum carpets against the dimensionless length and time for the case $\lambda_{c}<< L$ and $q=10^{-6}$. In this case, we take $m_{min}=1$, $m_{max}=20$ in Fig. 1(c) while $m_{min}=1$, $m_{max}=40$ in Fig. 1(d). It is clear from Figs. 1(c,d), considering higher quantum numbers more prominent quantum carpet can be seen as indicated in Fig. (d). It is noted that the revival time of the initial wave packet remains the same in Figs. 1(c,d).
\section{Spatio-temporal of KG particle} 
In this section, we study the spatio-temporal of KG particle of mass $M$ placed in a one -dimensional box. The mathematical expression for the KG particle in a one-dimensional box can be written as 
\begin{equation}\label{eq:kg}
(\Box +Mc^{2})\xi(z,t)=0.
\end{equation} 
Where $$\Box=\frac{1}{c^{2}}\frac{\partial^{2}}{\partial t^{2}}-\frac{\partial^{2}}{\partial z^{2}}.$$
We use the separation of variable technique i.e., $\xi(z,t)=f(z) g(t)$, the plan wave solution of the Klein Gordon equation within the box is
\begin{equation}\label{eq:sol2}
\begin{split}
f(x)=A\sin(kz)+B\cos(kz), \\g(t)=C\exp[\frac{iEt}{\hbar}]+D\exp[\frac{-iEt}{\hbar}],
\end{split}
\end{equation} 
the spatial part of Eq.~\eqref{eq:sol2}, under the boundary conditions $f(z)=0$ at $z=0$, and $z=L$, the eigenfunction can now be read as
\begin{equation}
f_{n}(z)=C^{\prime}_{n}\sin(\frac{n\pi z}{L}).
\end{equation}
The complete eigen function for the KG particle after time t in a one-dimensional box modifies 
\begin{equation}
\xi(z,t)=A^{\prime}_{n}\exp[\frac{iEt}{\hbar}]\sin(\frac{n\pi z}{L})+B^{\prime}_{n}\exp[\frac{-iEt}{\hbar}]\sin(\frac{n\pi z}{L}),
\end{equation}
where $A^{\prime}_{n}$ and $B^{\prime}_{n}$ are respectively normalization constants which can be evaluated through normalization conditions $\int_{-\infty}^{\infty}\rho^{\pm}d^{3}z=\pm e$. Here, $\rho^{\pm}=\hbar e/2iM(\xi^{\star \pm}\partial\xi^{\pm}/\partial t-\xi^{\pm}\partial\xi^{\star\pm}/\partial t)$. The time-dependent wave function of the KG particle is
\begin{equation}
\xi^{\pm}_{n}(z,t)=\sqrt{\frac{2M}{LE_{n}^{\pm}}}\sin(\frac{n\pi z}{L})\exp[\frac{iE_{n}^{\pm}t}{\hbar}],
\end{equation}
where $\pm$ signs indicate particle and anti particle corresponding to energy $E_{n}^{\pm}=\pm c\sqrt{\frac{n^{2}\pi^{2}\hbar^{2}}{L^{2}}+M^{2}c^{2}}$. The complete wave function of the KG particle at time $t=0$ is given by 
\begin{equation}\label{eq:timez}
\begin{split}
\chi_{n}(z,0)=&\frac{1}{\sqrt{2\pi\Delta}}\sum_{n=1}^{\infty}\sqrt{\frac{2M}{LE_{n}}}\exp[\frac{-k_{n}^{2}\Delta^{2}}{2}]\sin(\frac{n\pi z}{L})\times\\&\sin(\frac{n\pi z_{0}}{L}).
\end{split}
\end{equation}
\begin{figure}[ht]
\centering
\includegraphics[height=6.5cm,width=4cm]{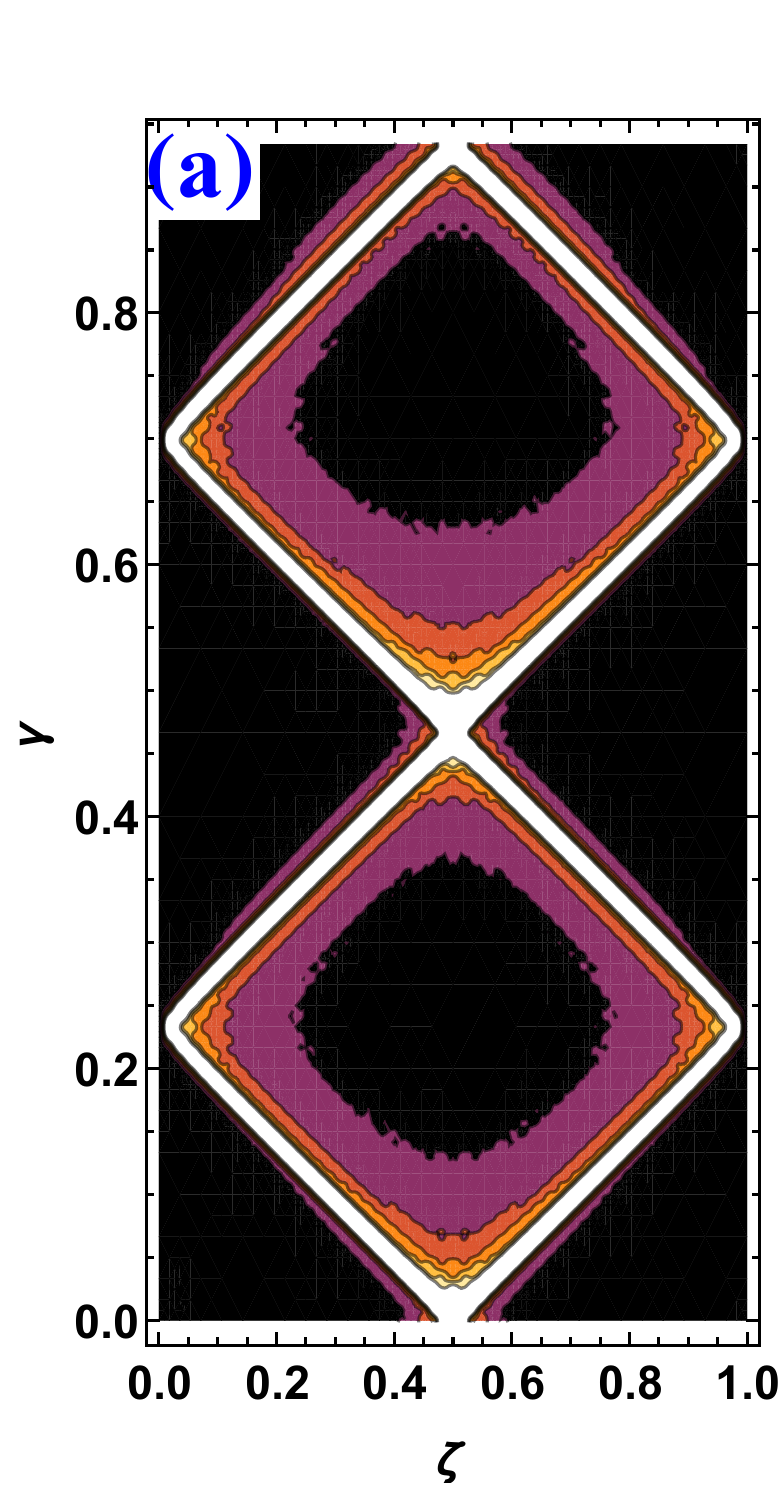}
\includegraphics[height=6.0cm,width=4cm]{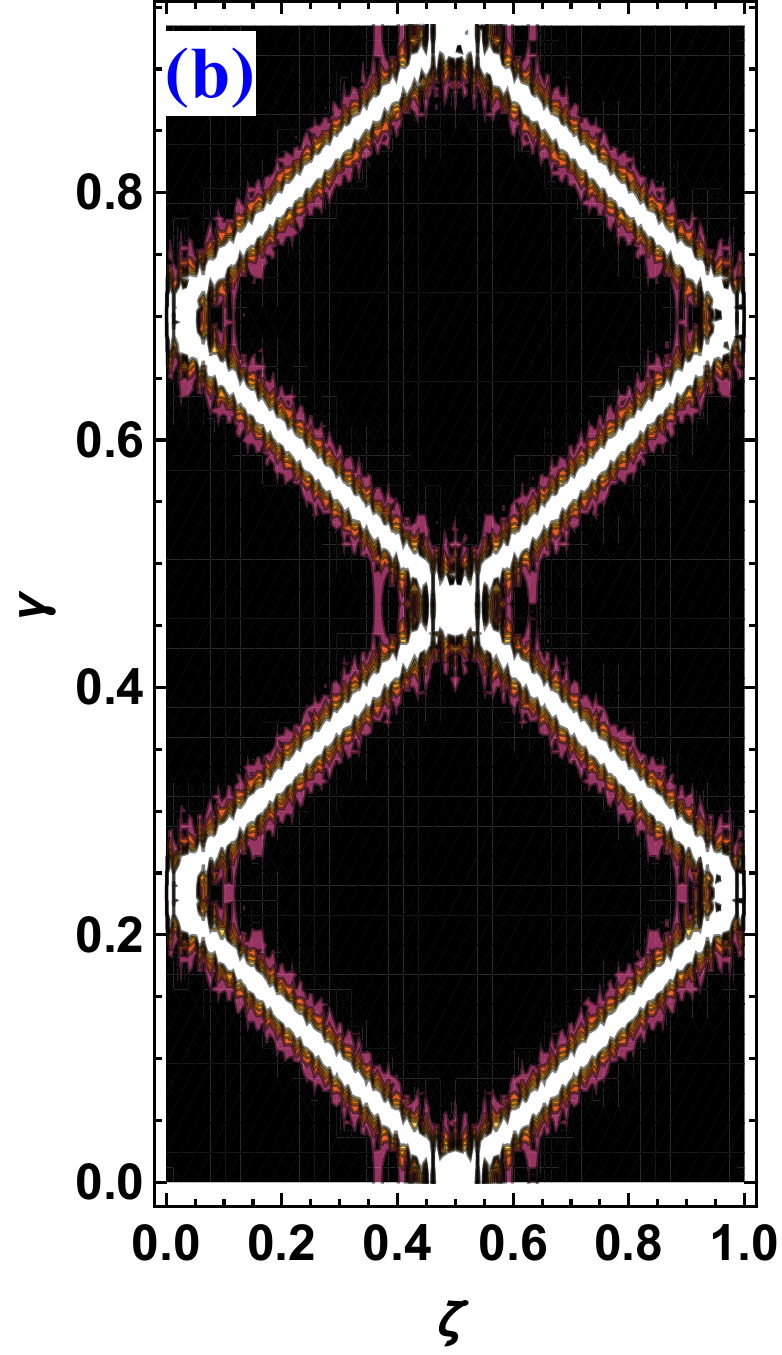}
\caption{The quantum carpets for the KG particle in a one-dimensional box is shown for the cases (a) $m_{min}=1$, $m_{max}=50$ and (b) $m_{min}=65$, $m_{max}=120$. Here, the scaled axes represent dimensionless length and time i.e., $\zeta$=$z/L$ and $\gamma=t/T^{rev}_{KG}$, respectively. We take the width of the Gaussian wave packet in order of $\Delta=10^{-2}$ in (a) and (b).}
\end{figure}
For space-time dynamics, we introduce the time evolution operator $\exp[\frac{-iE_{n}t}{\hbar}]$ in Eq.~\eqref{eq:timez}, the wave function evolves with time t is 
\begin{equation}
\begin{split}
\chi_{n}(z
,t)=&\frac{1}{\sqrt{2\pi \Delta}}\sum_{n=1}^{\infty}\sqrt{\frac{2M}{LE_{n}}}\exp[\frac{-k_{n}^{2}\Delta^{2}}{2}]\sin(\frac{n\pi z}{L})\times\\&\sin(\frac{n\pi z_{0}}{L})\exp[\frac{-iE_{n}t}{\hbar}].
\end{split}
\end{equation}
The probability of finding of the KG particle at position z and at time t can be evaluated as
\begin{equation}
\begin{split}
P(z,t)=&M/\pi\Delta L\sum_{n=1}^{\infty}\sum_{m=1}^{\infty}\frac{1}{\sqrt{E_{n}E_{m}}}\exp[-\frac{\pi^{2}\Delta^{2}}{2L^{2}}(m^{2}+n^{2})]\\&\times\sin(\frac{n\pi z}{L})\sin(\frac{m\pi z}{L})\sin(\frac{n\pi z_{0}}{L})\sin(\frac{m\pi z_{0}}{L})\times\\&\exp[\frac{-i(E_{n}-E_{m})t}{\hbar}].
\end{split}
\end{equation} 

We plot the probability density P(z,t) in Figs. 2(a,b) versus scaled parameter $z/L$ and $t/T^{rev}_{KG}$ by considering the quantum numbers from lower to upper values, $m_{min}=1$, $m_{max}=50$ in (a) and $m_{min}=65$, $m_{max}=160$ in (b) respectively. One can see from Figs. 2(a,b) considering the lower quantum numbers much denser is the quantum carpets of the bosonic particle and vice versa. This means that the density of the quantum carpets of a bosonic particle trapped in a box has a inverse relation with the quantum numbers. In addition, it is clear from Figs. 2(a,b), the bosonic particle reconstruct itself in a specific time $t=T^{rev}_{KG}$ and $T^{rev}_{KG}=\pi\hbar\sqrt{1+4n_{0}^{2}q^{{\prime}^{2}}}/(2Mc^{2}q^{{\prime}^{2}}n_{0})$ with $q{\prime}=1$, which remain fixed in both cases.\\
Examine the revival time of the Dirac particle with KG particle. We found that $T^{rev}_{Dirac}>T^{rev}_{KG}$, the reason is that each particle has a different mass. The bosonic particle is much heavier than fermion, in other words, the energy of the bosonic particle is greater than the fermion. As a result; the revival time of the bosonic particle is smaller than fermionic particle. It means that, if we have two different particles (boson and fermion) and are separately confined in a one dimensional boxes of same height. Suppose each particle evolves with the same time, the bosonic particle will reconstruct itself earlier than fermionic particle due to its shorter revival time. 
\section{conclusion}
The spatio-temporal of relativistic particles confined in a one dimensional box has been discussed. First, we have considered a fermionic particle in a one dimensional box and studied the quantum carpets at relativistic regime. Further, we imposed different limiting conditions on Dirac energy solution and explained slight-relativistic and non-relativistic quantum carpets, respectively. Next, we have introduced a bosonic particle within the box and a detailed analysis of the quantum carpets and revivals have been delivered. Furthermore, we have discussed different revival times for fermionic particle and bosonic particle and found that $T^{rev}_{Dirac}>T^{rev}_{KG}$, in a one dimensional box confined with the same height. In addition, we showed that the density of the quantum carpets depend on the corresponding quantum numbers. We observed that the quantum carpets of the bosonic particle is much denser than fermionic quantum carpets, for lower quantum numbers, but less denser, for upper quantum numbers.

\section*{Acknowledgment}

The authors would like to thank the anonymous referee for his useful comments which greately improved our manuscript. This work was supported by the department of Physics, Quaid-i-Azam University Islamabad.


\end{document}